# The Case for API Communicability Evaluation: Introducing API-SI with Examples from Keras

Luiz Marques Afonso[1], João Antonio Marcondes Dutra Bastos[1],
Clarisse Sieckenius de Souza[1] and Renato Fontoura de Gusmão Cerqueira[1,2]

Semiotic Engineering Research Group (SERG) [1]
Departamento de Informática, PUC-Rio
{lafonso, jbastos,clarisse}@inf.puc-rio.br
Rua Marquês de São Vicente, 225 – Rio de Janeiro, RJ – Brazil

IBM Research Brazil [2]
Natural Resources Solutions
rcerq@br.ibm.com
Avenida Pasteur, 138 – Rio de Janeiro, RJ – Brazil

*Abstract*—In addition to their vital role in professional software development, Application Programming Interfaces (APIs) are now increasingly used by non-professional programmers, including end users, scientists and experts from other domains. Therefore, good APIs must meet old and new user requirements. Most of the research on API evaluation and design derives from user-centered, cognitive perspectives on human-computer interaction. As an alternative, we present a lower-threshold variant of a previously proposed semiotic API evaluation tool. We illustrate the procedures and power of this variant, called API Signification Inspection (API-SI), with Keras, a Deep Learning API. The illustration also shows how the method can complement and fertilize API usability studies. Additionally, API-SI is packaged as an introductory semiotic tool that API designers and researchers can use to evaluate the communication of design intent and product rationale to other programmers through implicit and explicit signs thereof, encountered in the API structure, behavior and documentation.

## I. INTRODUCTION

Interest in the quality of Application Programming Interfaces (APIs) design has been growing steadily in recent years. A key factor is that non-professional programmers, like end user developers, scientists and experts from other domains [1] have been using APIs to produce software that they build by themselves. Usability has thus become an especially critical requirement for APIs. Research on API usability harnesses a wealth of knowledge produced in the field of human-computer interaction (HCI). It has been helping API designers to understand their professional and non-professional users' needs and to deliver software packages that are more likely to improve such users' effectiveness, efficiency and satisfaction with software development activities.

Most of the research to-date – as suggested by the expression '*API usability*' – is strongly influenced by classical user-centered design views [2]. Human factors of interest, in this perspective, are all referenced to *users*. Therefore, the study of such factors provides usability goals that the designers of interactive technologies can meet with knowledge, techniques and tools provided by user-centered research.

Starting with the work of Andersen [3] and Nadin [4] in the 1980's, the often-called *semiotic approaches to HCI* have since their inception taken a different perspective. Semioticians who study human-computer interaction see it as a special case of computer-mediated human communication, engaging systems designers (software producers) and systems users (software consumers) in a kind of *conversation* carried out using interface languages, programming languages, markup languages, or other computer language. This more evenly balanced view of the designers' and users' participation in HCI, as conversational *partners*, has generated a substantial volume of semiotic theory and methods, much of it associated with Semiotic Engineering [5] [6].

The first international publication proposing a Semiotic Engineering evaluation tool for APIs appeared in 2016. Our own *SigniFYIng APIs*, as we called it, is one of several modules in the *SigniFYI Suite* of tools for the investigation of meanings and signification in human-centered software development [7]. However, *SigniFYIng APIs* can be challenging for those who are not familiar with the semiotic way of seeing things. Hence there is room to create simpler variants and scaffolds of *SigniFYIng APIs* to help interested researchers expand their knowledge, perception and skills. This is thus our goal, to support the appropriation and effective use of Semiotic Engineering concepts and techniques by API designers and researchers. They should thus be able to evaluate the communication of design intent and product rationale to other programmers through implicit and explicit signs thereof, encountered in the API structure, behavior and documentation.

This paper presents a simpler semiotic evaluation method called *API Signification Inspection* (API-SI). It is a lower-threshold and more scaffolded variant of *SigniFYIng APIs* [7]. API-SI is packed as a technical tool and illustrated with highlights from the semiotic analysis of *Keras* [8], a popular API in the Deep Learning (DL) domain. *Keras* gives us a prime opportunity to see how Semiotic Engineering, in general, and API-SI, in particular, articulate the knowledge that API designers have about *themselves* and *their own activity* in order to improve the communicability of application programming interfaces. In 2017 François Chollet, the creator of *Keras*, published a book with a rich description of his vision, design intent, and product rationale [9]. As a faithful representation of what the API designer *means to communicate*, the content of the book is then used in this paper to show how well the structure, the features and the behavior of *Keras* effectively *communicate* Chollet's message to this API's users.

API-SI touches the essence of the *communicability evaluation* of user interfaces in general [6] [7] and brings the Semiotic Engineering perspective closer to the community who is specifically interested in the design, evaluation and use of application programming interfaces. With selected illustrations centered on the problem of *overfitting* neural network models, we show how the technical scaffolds and analytic procedures of API-SI promote intense reflective and epistemic activity. The more systemic results of semiotic analysis, compared with some usability evaluation techniques, enable API-SI adopters to see potentially unsuspected connections between personal values, beliefs and intentions, on the one hand, and technical design choices about API features, structures and behavior, on the other. Moreover, when used by API designers themselves, our method should enable them to explore these connections in richly informed ways and potentially make better or more creative design decisions. Communicability and usability evaluations thus take different paths but ultimately share the same destination – both seek to help API designers to improve the quality of their work and make APIs more accessible to increasingly diverse communities of users.

In the next sections we: review the related work (section II); present the logic, the analytic scaffolds, and the procedures of API-SI (section III); highlight and discuss noteworthy results of the analysis of *Keras* produced with API-SI (section IV); and finally conclude the paper with suggestions for future work (section V).

## II. Related Work

API usability has brought together research communities that haven't often collaborated in the past. In 2016 Myers and Stylos [10] have published a widely circulated article calling API designers' attention to large volume of knowledge that they can use to meet their users' needs and, as a result, decrease the number of errors that programmers make when using APIs designed by somebody else. The authors mention evaluation tools, like Heuristic Evaluation [11] and the Cognitive Dimensions of Notations (CDN) framework [12], which have been originally proposed for HCI but can help assess the usability of API designs and inform more usable design alternatives. CDN has been extensively used in pioneering research about the quality of APIs [13] [14], but so have other HCI methods like classical user studies (e.g. [15], [16], [17]). Closer to software engineering perspectives, some approaches have explored automatic API evaluation techniques (e.g. [18], [19]). Historically, however, the Natural Programming Project [20] [21] [22] and the Psychology of Programming Interest Group – a research community strongly committed with the developments of CDN [23] [24] [25] – are long-standing contributors to the study of what is now mostly called a "human-centered" perspective on programming [26], which in turn has been stimulating considerable advances in API usability.

Since our work centers on inspection methods, we acknowledge approaches published in recent years. Farooq and Zirkler proposed a group-based method to evaluate the usability of APIs [27], where peers inspect the API. O'Callaghan [28] proposed a walkthrough method to evaluate APIs, where a participant, observed by the evaluator, walks through a prototype of the API ("a static text document containing a series of programmatic statements"). Both Farooq and Zirkler's and O'Callaghan's methods have the advantage of getting the actual reactions of potential API users with the inspection. A stronger method is Grill, Polacek, and Tscheligi's methodology that combines widely used HCI methods and techniques (heuristic evaluation, developers workshop and interviews) to evaluate the usability of APIs [29]. In 2017, Wijayarathna, Arachchilage, and Slay [30] proposed a CDN-based questionnaire to evaluate security in APIs. The work is inspired in Clarke's early work, but focuses more narrowly on aspects of API design that expose programmers to vulnerability issues.

To the best of our knowledge, the most comprehensive set of usability heuristics and guidelines for APIs has been published in 2018, by Mosqueira-Rey and co-authors [31]. They present a thoroughly detailed review of the literature in the area and conclude that most of the research on API usability to date is "very technically-minded and tends to neglect the subjective component of usability". They organize the findings and proposals of previous research into two taxonomies (API usability and context of use), from which they derive – by cross-combination of elements – their set of heuristics and guidelines.

Compared to the related work, ours shares the emphasis on *notations* with CDN, inasmuch as all semiotic approaches take *signs* (i.e. *representations*) as the fundamental unit of analysis. Our perspective, however, is remarkably different from cognitive dimensions and, as the next sections will show, from *usability* itself. Our focus is placed on *communicability* and the two groups of humans involved in it, API users **and** API designers. We also share, as mentioned earlier, the types of procedures used in most inspection or walkthrough methods. However, at least at this stage, we do not require that our method's inspection be made in group or with the participation of an API user representative (which is always an option for the evaluator). Last, but not least, we find it revealing that Mosqueira-Rey and co-authors [31] translate the API acronym as "application program interface", rather than as "application programm*ing* interface", and that they define APIs as "a particular set of rules and specifications that software programs can follow in order to communicate with *each other*." (p. 46, our emphasis) Our method brings the humans behind and before software programs to the front stage and, instead of tacitly anthropomorphizing programs, openly views programs (APIs) as *proxies* who speak for their creators and put them in interface-mediated communication with API users, during application *programming* activities.

III. INSPECTING SIGNIFICATION IN APIS WITH API-SI

Like most inspection methods (*e.g.* Heuristic Evaluation [11], Cognitive Walkthrough [32], the Cognitive Dimensions of Notations framework [12]), API-SI is a *qualitative method*. "Qualitative methods usually generate data not easily reduced to numbers, [whereas] quantitative methods result in data to which the power of mathematical analysis can be applied easily" [33]. The main analytic activity when working with qualitative methodology is to assign and elaborate meanings associated to the data, to recognize the contexts and variations in which they occur, and finally to derive a collection of consistently related higher-level meaning categories th at can, typically but not exclusively, drive the inferences required to answer 'how' questions about the data. [34]

| Intentional Software Production and Use | Signifying Human Intent in Software Code and Interaction |
|---|---|
| (1) Intent (or intention) is more than a target or a goal; it involves a complex collection of articulated values, beliefs and expectations regarding oneself, others and the world. | (5) The producers' and the consumers' intent are manifested and achieved through communication in *unique* computable interface languages, that is, intentionally designed and formally specifiable collections of symbols that can be interpreted and/or generated by computer programs. |
| (2) Anyone who writes a computer program does it with an intent in mind. | (6) Interface languages are *unique* because their vocabulary and semantics necessarily express the producers' *unique* intent regarding the program. |
| (3) Anyone who uses a computer program does it with an intent in mind. | (7) Users must learn *unique* interface languages to express their own intent, and they do so by being exposed to them and using them, similarly to learning a new natural language. |
| (4) A computer program is a medium where its producers' and consumers' intent meet and are transformed into action through user interfaces. | (8) Patterns of language syntax and semantic interpretation can be accentuated and exploited by software designers to *teach* software users how to communicate their intent using new interface languages. |

**Computer-Mediated Communication for the Achievement of Programmers' and Users' Intent**

(9) The success of encounters between producers' and consumers' intent through computer interfaces depends heavily, though not exclusively, on how well producers and consumers can: manifest their intent; understand each other's response to such manifestations; present, explain and negotiate conditions and alternatives; and thus, achieve mutual satisfaction.

(10) Although computer programs interpret and generate interaction algorithmically, humans are virtually incapable of doing so. Every time they use the same interface language expression, users gain new subjective and objective experience with it, and accumulate new conscious or unconscious knowledge.

**Figure 1: The Semiotic Engineering Factsheet**

*A. API-SI Logic and Analytic Scaffolds*

API-SI, as its predecessor *SigniFYIng APIs* [7], analyzes *how* design intent and rationale is communicated to programmers (the API *users*) through an API's vocabulary, structure, run-time behavior, and documentation. As a semiotic method, it focuses on how the API designers' beliefs, intent, values and resources are explored and combined to meet not only the needs and expectations of the API users, but also those of the API designers, themselves. The underlying assumptions are that: API designers want and need to satisfy the users; API designers also want and need to satisfy goals, constraints and criteria that may not be directly related to users; and finally, good API design is the result of making the best choices and tradeoffs while trying to satisfy both parties' expectations and needs. These assumptions permeate the main tenets of Semiotic Engineering [5], which are presented in the *Semiotic Engineering Factsheet* (Figure 1).

From (1) to (4), the statements in the factsheet underline that: intent (or intentions) are more than just goals or purposes; both software producers and software consumers act intentionally; and finally that the producers' and consumers' intent come together when software is used. From (5) to (8), the statements underline that software code and behavior (which includes execution and supported interactions) express, or signify, the producers' and consumers' intent in a *unique* interface language. The language is unique because each software producer's intent is *unique* (cf. the definition for intent in (1)). Therefore, even if there are strong similarities with other languages, there are inevitably semantic specificities (*meanings*) that are different from those of every other interface language (lest two pieces of software are identical to one another). Finally, statements (9) and (10) underline fundamentally important conditions for the success of computer-mediated human communication through interface languages: that interlocutors who take part in communication must be able to explain and negotiate meanings through conversation; and that computer programs (who *speak* for their producers at use time) and humans do not interpret and generate meanings in the same way.

Like our previously proposed semiotic tools to support API design and evaluation [35] [7], API-SI is a strongly epistemic tool. It promotes and sustains in-depth reflection about the API designer's intent, activity, perception of API users, and relation with them. It therefore adds an interpersonal, social dimension to API design that user-centered and cognitively oriented usability evaluation tools do not address in comparable terms (if at all). This focus on computer-mediated social interaction between software producers and software consumers opens the door to the analysis and design of a separate pragmatic component of computer language descriptions. Programming, specification, modeling, querying and even many interaction languages are traditionally defined only in terms of vocabulary, syntax and semantics. Pragmatic elements, when considered, are usually part of the semantic component. Ideally separable from other language components, a pragmatic component should specify at least the following:

| Beliefs about *users* | Beliefs about *the API* | Beliefs about *the role of communication in the quality of the API users' experience* |
|---|---|---|
| • Who is the TARGETED AUDIENCE ([TA])? | • Has this API been especially designed for the [TA]?<br>　o Does the API contemplate their values? | • Where do you say who your [TA] is and what features can best demonstrate that the API has been *made for them*?<br>• Can non-targeted audiences easily realize that the API *is not for them*?<br>　o Where do you say it? |
| • What does the [TA] know? | • Does the [TA] already know a substantial part of what is needed to benefit from using the API?<br>• How much new knowledge does the [TA] need to learn?<br>　o Will the [TA] be able and willing to acquire this new knowledge? | • How does the API vocabulary, syntax, semantics and run-time behavior communicate what you expect the [TA] to know already?<br>• How do the documentation content and access modes, run-time messages (including warnings, error messages, and other kinds of externalized information), as well as user-API interaction help the [TA] acquire and retain new knowledge? |
| • What does the [TA] want/need to do? | • What in-built functions does the API provide to the [TA]?<br>• If the [TA] needs new functions, can the API be extended?<br>• If the [TA] wants to change an in-built function, can the API be changed? | • How can the [TA] know what functions are available?<br>• How do you support the [TA] in deciding whether they can or should add a new function to the API?<br>• How do you tell the [TA] that they can or cannot change the API's in-built functions?<br>　o If in-built functions can be changed, how do you tell the [TA] that they can or cannot restore the original API version? |
| • Why? | • How do the in-built functions match the anticipated practical programming needs of the [TA]?<br>• What are the plausible reasons why the [TA] would need or wish to:<br>　o Add new functions to the API; or<br>　o Modify in-built functions? | • Do you tell the [TA] what are the reasons for your API design choices?<br>　o How do you do it? |
| • Where and when? | • What are the physical and conceptual contexts and conditions in which you expect the [TA] to use the API? | • How does the API vocabulary, syntax, semantics and run-time behavior communicate the contexts and conditions in which you expect the [TA] to use the API? |
| • How? | • How can the [TA] invoke, understand, explore, test, debug and enhance the use of the API while:<br>　o Building software *in isolation* from the larger programming context where the API functionality is needed; and<br>　o Building software *in integrated mode*, where the API is just part of a larger programming structure? | • How do the documentation content and access modes, run-time messages (including warnings, error messages, and other kinds of externalized information), as well as user-API interaction help the [TA] know the contexts and conditions in which you expect the API to be used?<br>• What programming features – including code, run-time, easily accessible help, and sensitivity to different contexts and conditions of software development – translate your beliefs about the [TA]'s:<br>　o Needs;<br>　o Expectations;<br>　o Preferences;<br>　o Values; and<br>　o Learning opportunities? |

**Figure 2: API-SI Analytic Guide**

- How the context of language use affects and changes the semantic meaning of linguistic objects;
- What mechanisms can be used in communication settings to understand and negotiate mutual meanings; and
- How distinct types of conversations (*e.g.* conversation for straight execution, for clarification, for guidance, for error detection, for error correction, for exploration, for simulation, etc.) can be started, structured and controlled.

In Figure 2 we show the API-SI Analytic Guide. With it API designers or evaluators can reason about what the designers think they are doing, for whom, why, how, etc. Note the strong connection and redundancy among items listed in rows and columns of the table in Figure 2, and also how these items relate to the Semiotic Engineering Factsheet in Figure 1. Note also that when analyzing somebody else's *intent*, while evaluating another designer's API for example, an evaluator behaves exactly in the same way as humans do in natural social settings. Using cultural knowledge, common sense, logic reasoning and interpretation of message cues and context, the evaluator – acting as the targeted users' advocate – constructs semiotically-justifiable assumptions about the designer's intent.

The guide is structured in three blocks (table columns). The first refers to the designer's beliefs about: who the users are;

what they know; what they want/need to do; why; where; when; and how. The second refers to the designers' beliefs about the API itself, reflecting not only how well the API fits the targeted users' profile, but also how well it achieves the designers' own intent and manifests the values that they want their product to have. Finally, the third brings users and designers together into the context of computer-mediated communication, helping the evaluators explore how the quality of *messages* that can be directly or indirectly communicated by one party or the other, through the *application programming interface*, affects the users' experience with the API. All three blocks contain questions or topics that the evaluator must consider. The tabular representation shows how questions and topics relate to others in the same row or column.

| Communication Channels | Communication Qualities |
|---|---|
| *Online Help and Documentation*: Documentation on demand; search and browse mechanisms; warning and information dialogs; embedded information at run-time. | *Consistency*: Messages communicated in all channels are consistent with each other. |
| | *Completeness*: The sum of messages communicated in all channels covers all that API designers want or need to tell the users. |
| *Coding Vocabulary, Syntax and Semantics*: lexical, syntactic and semantic patterns; analogies or conformity with known languages. | *Redundancy and Distribution*: Critical messages are communicated in multiple (possibly all) channels and multiple ways. |
| *Run-Time Behavior*: sensitivity to different contexts of execution (controlled by user or not); support to error prevention, detection, diagnose, and repair; user-configurable output and logging mechanisms. | *Meaning-Making Support*: The communication of computer language semantics and interpretive power can reach users by means of: consistent error prevention patterns; informative error handling (including detection, diagnosis and repair); incentive and support for experimentation and querying; and semiotically-rich behavior externalizations (potentially using different interface languages, focusing on different contexts and topics, allowing for different kinds of interactions). |
| *Interactive Means and Modes*: conversational support at coding time and run time; alternative interface representations or languages. | *Pragmatic Adequacy*: API users can choose communication means and modes, according to their own intent and preferences, as well as with the context and conditions of their activity. |

**Figure 3: Communicability Features Chart**

Together, the Semiotic Engineering factsheet (Figure 1) and the API-SI Analytic Guide (Figure 2) help API designers and evaluators understand the elements and the consequences of a semiotically-informed construction of static and dynamic representations of: what the API is; whom it has been designed to benefit, in which ways, contexts, conditions, and why; how the API works; what kinds of information it can provide about itself; and what kinds of interactions API users can have with it, when, where, how, for what purposes and why. In Figure 3 we show the communication channels that API designers can explore in this process, as well as communication qualities that contribute to making the signs in API code, behavior, documentation and support speak for the person (or team) who has designed it, effectively playing the role of the designer's *proxy* in mediated communication with API users.

Communication channels listed in the Communicability Features Chart (Figure 3) correspond to the standard sources of communication and expression that API designers can explore: online help and documentation; the coding language (vocabulary, syntax and semantics); the API behavior at run-time; and the kinds of interactions that API users can have with it. Communication qualities, in turn, are drawn from interpersonal collaborative and mutually helpful communication for action. Communication must: be consistent; be complete; be sufficiently redundant and well distributed across different means and modes of communication; support meaning-making by users who are learning or probing how the API works; and finally allow users to shape the kinds of interactions they wish to have with the API.

*B. API-SI Procedures*

The inspection procedures of API-SI can be seen on Figure 4. There are five steps in the method. Except for the initial and the final steps, the remaining three steps are part of an iterated cycle, where the inspector is carefully examining alternative paths of interaction with the API and collecting empirical evidence for the conclusion of the analysis. Note that Figure 4 also shows which scaffolds are used in which steps.

The Semiotic Engineering **Factsheet** sets the keynote for the entire process, and is used in all steps. The Analytic **Guide** is used in the first and third steps, and the Communicability Features **Chart** is used only in the last step of the process.

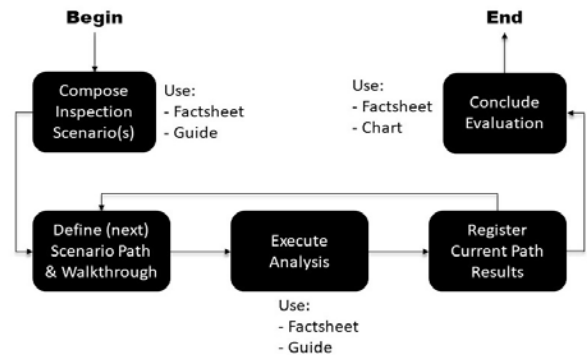

**Figure 4: The API-SI Procedures and Use of Scaffolds**

To **compose the inspection scenario**, the inspector will use the first column of the Analytic **Guide** and determine the profile of the targeted audience. The targeted user becomes the main character of the narrative that characterizes the scenario of inspection. Moreover, to ensure the evaluation of context-sensitive communication, the scenario must *focus* on some activity and provide sufficient information about the targeted user's situation. For illustration, we will focus the inspection on *Keras's communication about DL model overfitting*. Overfitting can be thought of as overgeneralization. The machine starts to learn patterns that are present in the data that is used to train it but misleading or irrelevant when it comes to new data [9]. A targeted user profile and his situation is defined in the example below.

*Martin is a graduate student taking an introductory course in Deep Learning. Joyce, his teacher, has just taught a lesson about model fitting and given the class a homework exercise with Keras. All students are sufficiently versed in Python, the programming language in which Keras is implemented. The exercise consists of comparing results when model parameters or model layers are changed. Students should learn how to recognize overfitting, a central concept in DL. The example used in the exercise is the well-known handwritten digit recognition problem using the MNIST dataset. Students should use Keras in the university's computer lab or install it in their own computers. They can freely use Keras books, online documentation, tutorials, and repository of examples to learn more.*

API-SI emphasizes the importance of exploring *alternative paths*. The inspection scenario is not a narrative to be used as a *specification of action*, but rather as a *richly characterized context for action*. During the core steps of analysis, the inspector will explore alternative paths within this context and see how the API responds to variations around the scenario's theme. To **start the analysis** the inspector chooses one path, for example 'to create a small and simple digit recognition model'. Martin might do this by copying and pasting an existing example from Keras documentation or online repositories.

To **carry on an entire cycle of analysis**, the inspector will walkthrough several possible ways how Martin might create a small and simple DL model with a copy & paste strategy. During the walkthrough, the inspector will bear in mind the content of the Semiotic Engineering Factsheet and examine what he sees in the light of issues and questions listed in the second and third columns of the API-SI Analytic Guide (Figure 2). For illustration, Martin might open Keras official documentation at https://keras.io/ and follow the link entitled "Getting Started: 30 seconds to Keras". In it, Keras designers are clearly and directly speaking to the users, as shown by the excerpt below.

"The core data structure of Keras is a model, a way to organize layers. The simplest type of model is the Sequential model, a linear stack of layers. […] Here is the sequential model […] Stacking layers is as easy as .**add()** […] Once your model looks good, configure its learning process with .**compile()** […] If you need to, you can further configure your optimizer. A core principle of Keras is to make things reasonably simple, while allowing the user to be fully in control when they need to (the ultimate control being the easy extensibility of the source code). […] You can now iterate on your training data in batches […] Evaluate your performance in one line […] Or generate predictions on new data […] Building a question answering system, an image classification model, a Neural Turing Machine, or any other model is just as fast. The ideas behind deep learning are simple, so why should their implementation be painful?"

The designer-user communication in the documentation is rich and eloquent. Martin should identify the digit recognition problem in the illustrated code on the page (marked by ellipsis in the excerpt above), since Joyce has used it in her slides.[1] At the end of this cycle of analysis, the inspector will see that by directly copying and pasting the code snippets from this particular documentation page and compiling the model, Martin will get the following error message: "line 15 in <module> model.fit(x_train, y_train, epochs=5, batch_size=32). NameError. Name x_train is not defined." This is because the content of the page does not explicitly mention another fundamental component for DL model building and training: the datasets that will be used. This may lead to the relevant communication breakdown captured by the inspector. The problem starts with documentation content and ends with run-time error message. The message that Martin should get from the sequence of events is that model compilation can only work if the code in Keras includes the necessary commands to load the training dataset. The inspector must therefore **register current results** of his analysis, with the corresponding issues and questions based on the Analytical Guide content, and proceed to the next cycle of analysis.

At the end of iterations, **to conclude the evaluation** the inspector will take all registered results and, by using the Semiotic Engineering Factsheet as support knowledge and the Communicability Feature Chart as reference, compose his final appreciation of the API communicability. On the left-hand side of the chart (Figure 3), the inspector will be prompted to look at *where* the signs of communication have been found (or not) and to the kinds of messages that they convey to API users. On the right-hand side, the inspector will be prompted to appreciate the style, resourcefulness, effectiveness and richness of communication strategies used in the API design. This is a major step in identifying improvement needs and opportunities in the API design, as well as in identifying well designed communication patterns and strategies that can be reused in the future.

To illustrate the kinds of conclusions that can be dawn with the chart, we can appreciate the designer's communication about data representation in Keras. When expressing his vision of Deep Learning and Keras [9], the designer tells us that the vast majority of machine-learning systems work with numerical data representations. (p. 31), and so does Keras. For example, with natural language processing, "like all other neural networks, deep-learning models don't take as input raw text: they only work with numeric tensors." (p. 180) Therefore, data representations must be *vectorized*. Keras has in-built utilities for encoding text. Programmers are encouraged to use them, since vectorization is a laborious and critical operation for machine-learning. This message is clearly communicated in documentation and tutorials, especially in Chollet's book. However, the message is *not* always well communicated in run-time behavior. A walkthrough of alternative paths will show that, if a large dataset with mostly numerical representations contains a segment of data unduly represented as characters, Keras silently processes the dataset *as if* all in it were numbers. This may create problems in the training process (how are characters interpreted in tensor manipulations?). These two communication strategies – to rely heavily on the attentive reading of documentation and to make benevolent interpretations of strictly malformed representations without bothering the programmer – can have costly effects in machine learning systems. The conclusion is that the communication of type checking rules in Keras should be better communicated and probably better controlled by the programmers, depending on the context and purpose of their work (a pragmatic decision).

---

[1] Note the selection of an alternative plausible walkthrough path (among several others), which is totally consistent with the scenario.

## IV. ILLUSTRATION AND DISCUSSION OF API-SI RESULTS

The results presented in this section follow the steps presented in Figure 4. The scenario of inspection is the same as section III. The inspection is thus guided by *Martin*'s profile and intent.

We start with the Keras designer's thoughts about model overfitting [9]. According to Chollet, a good DL model is one "that stands right at the border between underfitting and overfitting", but that "to figure out where this border lies, first you must cross it." (p.114). The proposed strategy is to begin by causing overfitting, by adding layers, making layers bigger and training for more epochs. Then, when the model overfits, "start regularizing and tuning the model, to get as close as possible to the ideal model that neither underfits nor overfits." (p. 114).

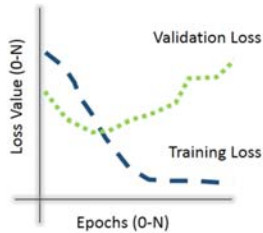

**Figure 5: Training x Validation Loss Data Plot Sketch**

In his book [9], overfitting conditions are communicated visually, with graphics generated by an external Python library (Matplotlib[2]). They effectively support the contrast between training and validation values for loss and accuracy. The underfitting/overfitting border can be easily identified in graphics like the one sketched in Figure 5 for the comparison of loss values. Visual communication of overfitting symptoms (or signs, in semiotic terminology) depends on the API user's initiative to use Matplotlib. We therefore ask how Keras – as a communication proxy for its designer – tells *Martin* that: (a) this can be done; and (b) this can help him understand the overfitting conditions and start eliminating them.

A search for "overfit" on the official documentation website for Keras (https://keras.io/) returns two pages[3], one about Noise Layers (two search string hits on this page, none referring to Matplotlib), and the other about Core Layers (two search string hits on this page, too, none referring to Matplotlib). The communication about overfitting in both cases focuses on how to mitigate it, but not on how to see it when it happens. Therefore, the official online documentation is not the preferred channel for communication about how *Martin* could detect overfitting. The inspector must look at other channels of communication.

The inspection of Keras's vocabulary, syntax and semantics leads to interesting findings. The documentation on "Callbacks" shows that there is a wealth of signs that communicate overfitting conditions. We highlight the most significant ones for the present illustration, EarlyStopping and ModelCheckPoint. EarlyStopping stops training when a monitored quantity (*e.g.* loss or accuracy) has stopped improving. One of this callback's arguments is patience, the number of epochs with no improvement after which training will be stopped. By default, the callback's *patience* (note the evocative name) is set to zero (no patience). Therefore, as soon as the monitored quantities are worse than they were in the previous epoch, the training stops. This is a strong communication of how critical the point of inflection in the validation loss data (Figure 5) is for Keras. Moreover, ModelCheckPoint saves the model after every epoch, which in combination with EarlyStopping allows the programmer to do exactly what Chollet expresses as his belief (and design intent) about model fitting [9]: work towards overfitting first; when overfitting is reached, start regularizing and tuning. Keras's run-time behavior provides evidence of communication that can be improved. For example, one way to call EarlyStopping is:

```
from keras.callbacks import EarlyStopping
early_stopping = EarlyStopping(monitor='val_loss', patience=2)
model.fit(x, y, validation_split=0.2, callbacks=[early_stopping])
```

If the programmer forgets to include the validation_split value in the model.fit() parameters, Keras does not have a default value for it. The run-time outcome is an error message like this:

```
callbacks.py:526: RuntimeWarning: Early stopping conditioned on metric
`val_loss` which is not available. Available metrics are:
acc,loss
```

Note that the error message contains signs of what the problem is ('val_loss' not available) and also about the solution ('Available metrics are: acc, loss'). However, the signs communicate the origin of the problem metonymically (*i. e.* using symptoms for the cause). The reason why loss and accuracy values are not available is because the programmer has not set apart a split of the dataset for validation. A more direct indication of the problem could be communicated with a simple addition of information, like: Early stopping conditioned on metric `val_loss` which is not available. <u>Check validation split</u>. Available metrics are: acc,loss. In this case, there is an explicit connection between metrics and the availability of data for computing it.

Run-time behavior also communicates additional kinds of 'actionable feedback', one of Keras's main design features (https://keras.io/why-use-keras/). Keras's fit() method is the heart of the training. "model.fit() returns a History object. This object has a member history, which is a dictionary containing data about everything that happened during training. [...] The dictionary contains four entries: one per metric that was being monitored during training and during validation." [9] One of fit() arguments is verbose, whose default value is set to printing out one line of information per epoch of training. Therefore, if the code in Keras contains a line like model.fit(data, labels, epochs=10, batch-size=32), the output on screen will start with following few lines:

```
Using TensorFlow backend.
Epoch 01/10
1000/1000 [=========] – 0S 259us/step – loss: 0.7076 – acc: 0.4960
Epoch 02/10
1000/1000 [=========] – 0S 38us/step – loss: 0.6952 – acc: 0.5110
...
```

Note that the listed lines include the information visualized in Figure 5. However, the textual printout is less efficient in communicating overfitting than the data plot, not only because of the merits of graphic communication in this case, but also because of other conditions. Unless the number of epochs is small, the user's screen will scroll a number of times, cutting off the initial training information (possibly the most important one) from the user's viewport. Run-time communication therefore does not do justice to the communication achieved with the vocabulary, syntax and semantics of the API in this case. Yet, this is not a matter of how much information is available to the user. It is a matter of how it is *presented*.

---

[2] https://matplotlib.org/

[3] Searches were run in March, 2018.

Finally, the inspection of **interactive means and modes** also leads to interesting findings. Interaction with Keras is split between coding time and run time. The user's (programmer's) communication is mainly expressed in the code (with different commands and parameters), whereas the designer's response is mainly expressed during execution[4]. This split makes designer-user *conversation* much harder than it would be if at least some conversational turns could be completely achieved at run time. For example, suppose that the following **pseudo-coded *conversation*** could take place:

Programmer at Coding Time: "Run program"
Designer at Run Time: "Here is the result; see progression of metrics"
Programmer at Run Time: "Run again with EarlyStopping, patience=5"
Designer at Run Time: "Here is the result; see progression of metrics"
Programmer at Run Time: "Run again with EarlyStopping, patience=0 and ModelCheckpoint"
Designer at Run Time: "Here is the result; Model saved with ModelCheckpoint default parameters. Type '?' if you want to see them or change them for the next run"
Programmer at Run Time: "?"
Designer at Run Time: "Here they are. Just type new parameter values if you want to: Run again with EarlyStopping, patience=0 and ModelCheckpoint."

This small snippet of pseudo-coded designer-user interaction for Keras illustrates what Semiotic Engineering researchers refer to as *conversational APIs* [36] and demonstrates some essential communicability features emphasized by API-SI scaffolds. As advised in the Semiotic Engineering Factsheet (Figure 1), we highlight the use of patterned features (item 8) and the possibility to manifest, explore and negotiate API meanings through interface languages and interaction mechanisms (item 9). We also highlight design features that contemplate what the targeted user wants to do and why, using the API-SI Analytic Guide (Figure 2) to bring interaction closer to facilitating communication about it (and action for it). Finally, we highlight once again the use of the Communicability Features Chart (Figure 3) to appreciate the current quality of Keras communication with programmers and to think of ways to improve it.

The above examples of Keras's communication of design intent, beliefs and values to users like *Martin* are inspected in the context of actions to understand and mitigate overfitting conditions in DL. The inspection suggests that, within the limits of our communicability evaluation's focus, Keras passes our consistency and completeness tests. However, we verify a few issues with other chartered communication qualities. For instance, with redundancy and distribution, a substantial part of information about overfitting is communicated only in online documentation and support, with reflexes in the API programming language. Yet, this communication is not well explored in run-time and interactive signs. Given the salience of the latter, redundancy and distribution of these messages over documentation *and* run-time channels is desirable. The pseudo-coded designer-user 'conversation', illustrated above, demonstrates how this might begin to be achieved.

Support to meaning making and pragmatic adequacy illustrate additional features of API-SI. Understanding and exploring overfitting conditions is, in Chollet's own words, one of the key factors in successful DL systems. Hence, the provision of efficient *signs* of overfitting conditions and sustained exploratory interactions around it can boost the power of Keras from "helping you win machine learning competitions" (https://keras.io/why-use-keras/) to *helping you master machine learning techniques and build solid applications of it* (our addition). Pragmatic adequacy should let users choose communication means and modes, not only according to their intent and preferences, but also to the context and conditions of their activity. For example, once the epistemic function of exploratory dialogs has been fulfilled, and the programmer understands overfitting, they are no longer needed. So, Keras should be used in *non-conversational mode* "to win machine-learning competitions". API-SI shows that Keras is consistently oriented towards competitions [9], but somehow misses the opportunity to play a powerful role in supporting scientific or professional development of programmers working with machine learning, with more epistemically-oriented communication.

V. CONCLUSION

This paper presented API-SI, a semiotic method to inspect how API designers' intent, beliefs, values and choices are communicated to API users through signs encountered in the API's multiple channels of communication. It is a more scaffolded and practical version of our own previously proposed *SigniFYIng APIs* [7]. We have shown that the semiotic approach is not detrimental to the users' interests, needs and expectations, emphasized in user-centered approaches. As demonstrated in section IV, API-SI allows us to detect both needs and opportunities to improve achieved design as well as to inform design alternatives that can be explored. The paper has also contrasted communicability with usability, two concepts that can greatly empower one another, and illustrated the concept of *conversational APIs*, which we have introduced in a previous publication [36].

Regarding future work, our long-term project is to use API-SI to improve the communicability of existing APIs and make them more conversational. A challenge that we anticipate is the need to accommodate two interaction modalities: a conversational style during more exploratory activities and a traditional style to support more straightforward programming. We also want to study communicability in API extension contexts, contrasting different contributors (API users or third parties) and scope (generic or domain-specific APIs). To this end, we want to study the pragmatic conditions (in the form of patterns or contracts) that developers of extension libraries should follow if they wish to ensure that their code will fit smoothly in the overall conversational features of the extended API. Likewise, we anticipate that extensions to support specific application domains should pay close attention to communicability and pragmatic aspects related to data manipulation and conversion.

---

[4] Some IDEs may offer improved interactivity at coding time.

We have just started a project with industrial partners to test the epistemic power of API-SI scaffolds in API design practice. The project will also give us the opportunity to *reify* conversational APIs, which we need to do in order to evaluate the impact of communicability on API *users*.

**Acknowledgments**

Clarisse S. de Souza and João A. M. D. Bastos thank CNPq, the Brazilian National Council for Scientific and Technological Development, for partially supporting this research (grants 304224/2017-0 and 142344/2016-8, respectively).